\documentclass[twocolumn,aps,pre,amsmath,amssymb,floatfix,longbibliography]{revtex4-1}
\usepackage{graphicx}
\usepackage{dcolumn}
\usepackage{bm}
\usepackage{amsfonts}
\usepackage{xcolor,tabu}
\usepackage{multirow}
\usepackage{amsthm}
\usepackage{tikz}
\usepackage[colorlinks=true,
            linkcolor=blue,
            urlcolor=blue,
            citecolor=blue]{hyperref}
\hypersetup{bookmarksopen=true}

\begin{document}

\title{Universal Scaling of Polygonal Desiccation Crack Patterns}

\author{Xiaolei Ma}
\email{xiaolei.ma@emory.edu}
\author{Janna Lowensohn}
\author{Justin C. Burton}

\affiliation{Department of Physics, Emory University, Atlanta, Georgia 30322, USA}

\date{\today}

\begin{abstract} 
Polygonal desiccation crack patterns are commonly observed in natural systems. Despite their quotidian nature, it is unclear whether similar crack patterns which span orders of magnitude in length scales share the same underlying physics. In thin films, the characteristic length of polygonal cracks is known to monotonically increase with the film thickness, however, existing theories that consider the mechanical, thermodynamic, hydrodynamic, and statistical properties of cracking often lead to contradictory predictions. Here we experimentally investigate polygonal cracks in drying suspensions of micron-sized particles by varying film thickness, boundary adhesion, packing fraction, and solvent. Although polygonal cracks were observed in most systems above a critical film thickness, in cornstarch-water mixtures, multi-scale crack patterns were observed due to two distinct desiccation mechanisms. Large-scale, primary polygons initially form due to capillary-induced film shrinkage, whereas small-scale, secondary polygons appear later due to the deswelling of the hygroscopic particles.  In addition, we find that the characteristic area of the polygonal cracks, $A_p$, obeys a universal power law, $A_p=\alpha h^{4/3}$, where $h$ is the film thickness. By quantitatively linking $\alpha$ with the material properties during crack formation, we provide a robust framework for understanding multi-scale polygonal crack patterns from microscopic to geologic scales.
\end{abstract}

\maketitle
\section{Introduction}
\begin{figure}[!]
\begin{center}
\includegraphics[width=3.4 in]{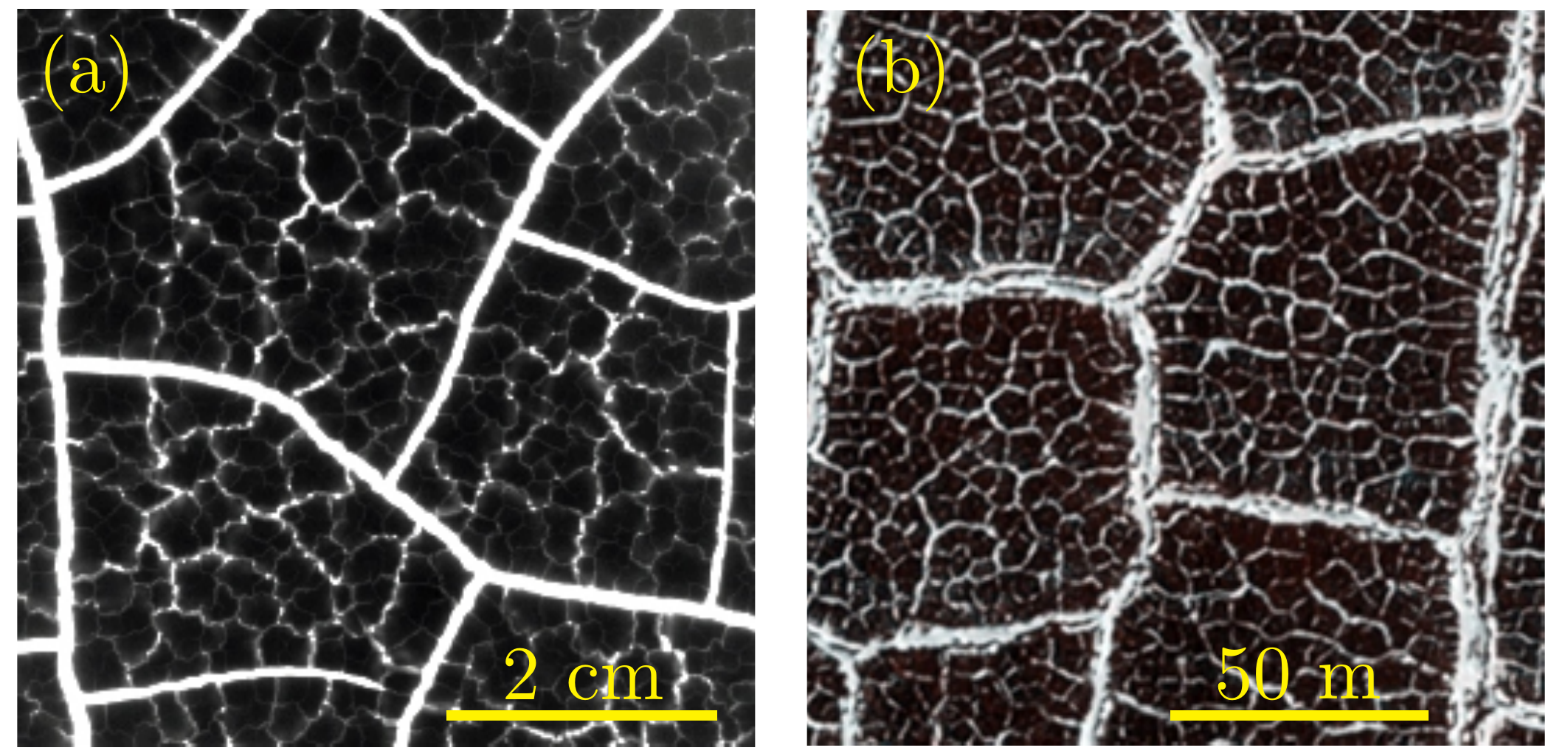}
\caption[]{Multi-scale polygonal crack patterns. (a) Crack patterns by drying cornstarch-water suspensions in a petri dish. (b) Polygonal terrain in an ancient dried lake be on Mars (HiRISE: PSP\_007372\_2475, image courtesy of NASA/JPL/University of Arizona). 
\label{multiscale_polygonal_cracks}
} 
\end{center} 
\end{figure}

Desiccation crack patterns observed in natural systems span many orders of magnitude in size (Fig.\ \ref{multiscale_polygonal_cracks}) \cite{ball1999self,xu2009drying,goehring2015desiccation,routh2013drying,goehring2013pattern,goehring2013evolving}. Among the large diversity of desiccation crack patterns, polygonal patterns are the most common. Typical examples include the complex crack network in dried blood \cite{brutin2011pattern}, craquelures in old paintings \cite{giorgiutti2015striped,goehring2017drying}, T/Y-shaped cracks in dried mud \cite{goehring2010evolution}, and polygonal terrain cracks \cite{goehring2014cracking,Maarry2010mars,el2014potential}. In particulate suspensions, the formation of desiccation cracks depends on the interplay between order and disorder in granular systems, as well as mechanical instabilities initiated by local, nonequilibrium interactions between the liquid, solid, and vapor phases \cite{routh2013drying,goehring2015desiccation}. Nevertheless, a broad range of practical applications, such as thin film coating, forensics, and controllable surface patterning rely on knowledge of the physical processes that determine crack patterns \cite{prosser2012avoiding,hatton2010assembly,liu2016surface,nam2012patterning,zeid2013influence}. Despite numerous studies which focus on desiccation crack patterns in a diverse range of systems, a fundamental understanding of the characteristic length scales associated with polygonal crack patterns is lacking, and it is not clear if the observed patterns in both microscopic and geologic crack patterns share the same underlying mechanisms.

In the laboratory, drying particulate suspensions, both Brownian and non-Brownian, are model systems for replicating and understanding desiccation cracks in nature. For a crack to form in any material, the mechanical potential energy released during fracture must exceed the energetic cost of creating new surfaces \cite{griffith1921vi}. In homogeneous, isotropic, elastic solids, the dynamics of fracturing have been recently characterized with exquisite detail   \cite{hutchinson1991mixed,fineberg1999instability,bouchbinder2010dynamics,bouchbinder2014dynamics,creton2016fracture}. However, the dynamics of drying-induced cracks are complicated by the lack of material homogeneity and a nonlinear relationship between stress and strain \cite{kitsunezaki2009crack,goehring2013plasticity,goehring2015desiccation,dufresne2006dynamics,lidon2014dynamics,man2008direct,xu2010imaging,xu2013imaging}. This complexity is enhanced by the multi-phase nature of the material: liquid menisci between particles generate heterogeneous shrinkage through capillary pressure, and excess liquid-vapor surface area in the bulk of the material \cite{dufresne2003flow,tirumkudulu2005cracking,singh2007cracking,chiu1993dryingI,chiu1993dryingII}. As a consequence, in addition to polygonal cracks, a large variety of cracks patterns have been reported in dried suspensions \cite{allain1995regular,inasawa2012self,goehring2011wavy,nandakishore2016crack,kiatkirakajorn2015formation,neda2002spiral,vermorel2010radial,darwich2012highly, jing2012formation,xu2010imaging,giorgiutti2015striped,giorgiutti2016painting,bohn2005hierarchicalI,bohn2005hierarchicalII,pauchard2003morphologies, boulogne2013annular,allen1987desiccation,gauthier2010shrinkage}. The variability in observed patterns depends on numerous factors such as film geometry \cite{lazarus2011craquelures,nandakishore2016crack}, particle mechanics \cite{nawaz2008effects}, liquid additives \cite{pauchard1999influence,liu2014tuning}, preparation history \cite{nakahara2006transition,nakahara2006imprinting}, solvent volatility \cite{boulogne2012effect,giorgiutti2014elapsed}, and external fields \cite{khatun2012electric,pauchard2008crack}.

Despite this broad range of crack patterns, we know surprisingly little about what controls the size and hierarchy of commonly-observed polygonal cracks, which are visible on both the micro and macro-scale. As shown in Fig.\ \ref{multiscale_polygonal_cracks}, a detailed understanding of desiccation crack patterns can lead to more accurate interpretations of planetary geomorphology, where data are limited to satellite-based imaging \cite{Maarry2010mars,el2014potential}. Thus far, laboratory experiments and numerical models have produced contradictory results as to the mechanism and dependence of the crack spacing on desiccation conditions. For example, for regularly-spaced cracks produced by directional drying, the relationship between the crack spacing, $\lambda$, and material thickness, $h$, is a power law, $\lambda\propto h^\beta$. Experimentally, numerous groups have reported $\beta\leq1$  \cite{allain1995regular,lee2004drying,smith2011effects}, and various theoretical predictions give $2/3\leq\beta\leq1$ \cite{ma2012possible,komatsu1997pattern}. For polygonal patterns, one may expect $A_p\propto \lambda^2\propto h^{2\beta}$. \citet{groisman1994experimental} reported $A_p\propto h^2$ in desiccated suspensions of coffee grinds, although only over a four-fold increase in $h$. Other experimental \cite{shorlin2000alumina} and numerical \cite{leung2000pattern,leung2010criticality} studies have reported similar scalings. Most recently, Flores \cite{flores2017mean} showed that $A_p\propto h^{4/3}$ using a model based solely on a balance of the average stress and surface energy released during cracking. Yet despite this history of investigation, there has been no systematic experimental study of the film's thickness and mechanics on the size of polygonal crack patterns.

The pattern morphology of polygonal cracks is also of interest since it reveals information about the formation and history of the cracking process \cite{goehring2015desiccation}. For example, the distribution of angles at crack junctions \cite{akiba2017morphometric,shorlin2000alumina,groisman1994experimental} and statistical analysis of the correlation length in crack patterns \cite{colina2000experimental} are commonly quantified from images of the surface. Repeated wetting and drying of the material can lead to more ``Y''-shaped junctions rather than ``T''-shaped junctions \cite{goehring2010evolution,goehring2014cracking}. For some commonly-used desiccation suspensions such as cornstarch-water  mixtures, crack patterns with two distinct length scales can be identified, as shown in Fig.\ \ref{multiscale_polygonal_cracks}a. For thick samples, the smaller polygons grow into the material, resembling columnar jointing patterns often found in nature \cite{muller1998starch,muller1998experimental,toramaru2004columnar,goehring2009drying,goehring2009nonequilibrium,akiba2017morphometric}. These smaller polygons are also known to coarsen with depth in the material \cite{goehring2005order}. Hierarchical patterns have also been observed in the cracking glaze of ceramics \cite{bohn2005hierarchicalI,bohn2005hierarchicalII}. It is not yet apparent why some materials display hierarchical crack patterns, and some do not. 

Here we present experimental evidence which resolves many of these important, outstanding questions. Our experiments involve analysis of desiccated crack patterns in various granular materials, such as cornstarch and CaCO$_3$, suspended in different volatile liquids: water, isopropanol (IPA), and silicone oil. We use both very thin, quasi-two-dimensional chambers, as well as open petri dishes to dry the samples. For all observed crack patterns, we find that the characteristic polygonal area is consistent with $A_p=\alpha h^{4/3}$ over more than three orders of magnitude in $h$, in agreement with a recent theoretical prediction \cite{flores2017mean} based on a balance of stress and surface energy for crack formation.  This scaling is independent of the shape of the polygons, which varies considerably depending on the material-liquid combination. By characterizing the modulus of the desiccated suspension, we are also able to quantitatively predict the prefactor $\alpha$. For all material-liquid combinations, we only observe multi-scale crack patterns in cornstarch-water mixtures. We show that these cracks are due to two distinct desiccation mechanisms. Primary cracks form first due to capillary-induced shrinkage of the material. Secondary cracks form much later, and are due to deswelling of the hygroscopic cornstarch particles. Taken together, these results provide a quantitative pathway for interpreting multi-scale polygonal desiccation crack patterns observed in diverse systems, from microscale colloidal films to terrestrial and extra-terrestrial planetary surfaces.

\section{Experiment}
We used commercial, polydisperse cornstarch particles from ARGO. The average radius, $R$, of the particles is $\approx$ 5 $\mu$m. We also used CaCO$_3$ particles from OnlineScienceMall with $R\approx$ 1 $\mu$m, and for some experiments, glass beads with $R \approx$ 5 $\mu$m from Miscrospheres-Nanospheres. Particle sizes were measured using optical microscopy. Different fluids such as deionized water, low-viscosity silicone oil (0.65 cSt, ClearCo), and 99.9\% isopropanol (IPA) were used as solvents to prepare particulate suspensions. For samples thicker than $h\approx$ 1 mm, we dried suspensions in polystyrene petri dishes of diameters 14 cm and 8.5 cm. Glass microscope slides were used to build thin, quasi-two-dimensional chambers as discussed in Sec.\ \ref{sec:secondary_cracks_thin_chamber}. For the thin chambers, the tunable thickness, $h$, was set by vinyl spacers cut from a sheet. Once the vinyl spacers were placed on the edges, a sample of suspension was placed on the microscope slide, and the sample was compressed by a second glass slide and secured mechanically before gluing with optical epoxy. Similar setups have been used by previous authors \cite{allain1995regular,dufresne2003flow,goehring2010solidification,inasawa2012self}. However, one important distinction in our experiments is that due to the relatively slow evaporation in this system, evaporation occurs nearly isotropically around the perimeter of the sample, so drying is not uni-directional. 

We used a conventional bright-field microscope to image crack patterns in the thin chambers, and a USB 3.0 digital video camera (Point Grey) connected to a macro lens to image the crack formation during drying of particulate suspensions in petri dishes from above. Recorded images were analyzed using NIH ImageJ software to obtain the area of polygonal cracks and thickness of the dried films. An electronic balance (Omega) was used to monitor the instantaneous mass of suspensions during drying. All experiments were performed at room temperature (20 $^\circ$C) with uncontrolled relative humidity of $\approx$ 60\%. Modulus measurements were obtained by slowly pressing a stainless steel ball of diameter 1.9 cm into the material using a rheometer (TA Instruments AR2000), and recording the applied normal force. 
\begin{figure*}[!]
\begin{center}
\includegraphics[width=6.5 in]{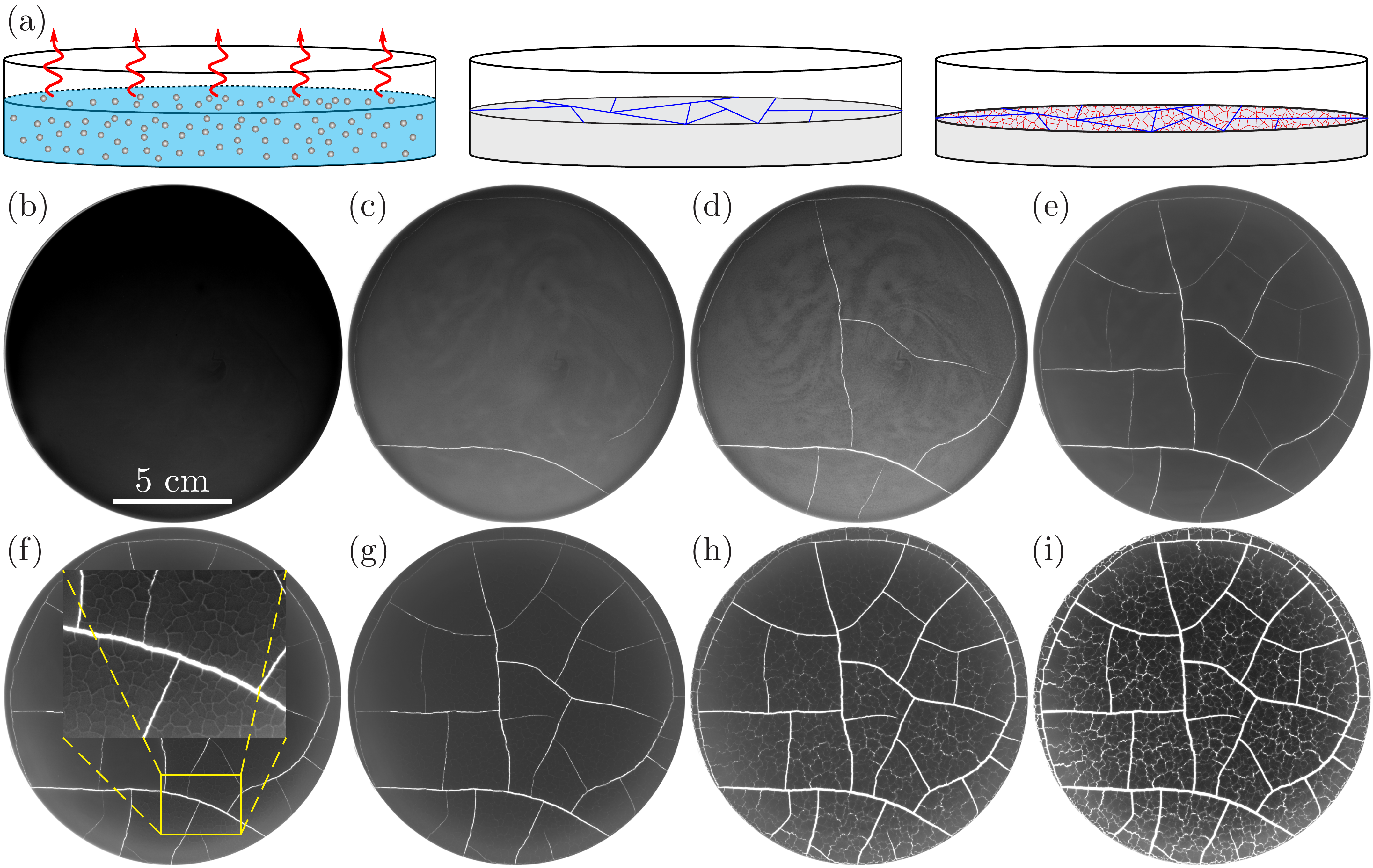}
\caption[]{(a) Experimental setup for drying cornstarch-water suspensions in petri dishes where drying occurs at the top interface as indicated by the red arrows. Formation of multi-scale crack patterns during drying of a cornstarch suspension ($\phi_i$ = 40\%) in a petri dish: (b) Image of the initial suspension ($t=0$), (c) the primary cracks appear ($t\approx$ 10.5 h), (d) the number of primary cracks increases ($t\approx$ 11 h), (e) the number of primary cracks stops growing ($t\approx$ 22 h), (f) secondary cracks appear ($t\approx$ 24 h), (g) the number of secondary cracks increases ($t\approx$ 29 h), (h) the number of secondary cracks stops increasing though drying still proceeds ($t\approx$ 42.5 h), (i) the final crack pattern ($t\approx$ 72 h). The inset in (f) is a zoomed-in view of the section enclosed by the yellow box. Some of the images here have been enhanced for the best visualization of crack patterns, and the scale bar applies to all images. The final dried film has a thickness of $h\approx$ 0.7 cm. A video (Video\_S1) showing the detailed formation of the multi-scale structures can be found in the Supplemental Material \cite{Ma2018cracks}.
\label{hierarchical_cracks}
}
\end{center} 
\end{figure*}

\section{Results and discussion}  
\subsection{Multiscale cracks in cornstarch-water suspensions}

We prepared cornstarch-water suspensions and dried the samples in petri dishes. We varied the thickness of the films by controlling the initial volume of the suspensions. Above a critical thickness, $h_c$, as will discussed in Sec.\ \ref{sec:crack_condition}, we observed two distinct crack patterns that appeared at different stages of desiccation. Figure\ \ref{hierarchical_cracks} shows the formation of these multi-scale cracks during drying of a cornstarch-water suspension with $\phi_i=40\%$. As shown in Fig.\ \ref{hierarchical_cracks}a, drying occurs at the air-water interface. The primary cracks (blue polygons) first appear, then as drying proceeds, secondary cracks (red polygons) appear within the larger polygons. After the suspension has dried for $t\approx$ 10.5 h, primary cracks first appear (Fig.\ \ref{hierarchical_cracks}c), and the number of primary cracks increases with time (Fig.\ \ref{hierarchical_cracks}d). At $t\approx$ 24 h, secondary cracks are visible (Fig.\ \ref{hierarchical_cracks}f). When $t\approx$ 42.5 h, the number of secondary cracks stops increasing though drying still proceeds (Fig.\ \ref{hierarchical_cracks}h). Finally, the drying contributes to the widening of the existing cracks, as shown in Fig.\ \ref{hierarchical_cracks}h. We note that the primary cracks penetrate completely through the sample when they form, whereas the secondary cracks grow more slowly, and their visibility increases with time. 

As reported by previous authors \cite{leung2000pattern,mizuguchi2005directional,akiba2017morphometric}, the primary cracks are a result of film shrinkage induced by the Laplace pressure on the scale of the particle size. As the water evaporates, menisci form between individual particles of radius $R\approx5$ $\mu$m. Thus the average pressure is reduced in the suspension by $\approx\gamma/R\approx$ 15 kPa, where $\gamma = 72$ mN/m is the surface tension of water.  Since the suspension is partially adhered to the bottom surface of the petri dish, the cracks form almost uniformly over the sample. Without this adhesion, the suspension undergoes isotropic shrinkage, and the number of primary cracks is reduced \cite{groisman1994experimental}. Figure S1 \cite{Ma2018cracks} shows the effects of different substrate boundary modifications on the primary crack patterns in cornstarch-water suspensions \cite{Ma2018cracks}. The secondary cracks are generally unaffected by the choice of boundary condition, and the origin of secondary cracks will be discussed in Sec.\ \ref{sec:particle_deswelling}.

\subsection{Critical condition for primary cracks}
\label{sec:crack_condition}
The appearance of primary cracks for thicker samples of cornstarch and water suspensions, as shown in Fig.\ \ref{hierarchical_cracks}, can be understood in terms of a well-known theory for the initiation of cracks in colloidal thin films \cite{tirumkudulu2005cracking,singh2007cracking}. The theory assumes a no-slip boundary condition between the bottom boundary of the film and the substrate, and predicts a relationship between the critical film thickness and stress when crack should appear:
\begin{equation}
\frac{\sigma_c R}{2 \gamma}=0.1877\left( \frac{2R}{h_c} \right)^{2/3}\left( \frac{G M \phi R}{2\gamma} \right)^{1/3},
\label{critical_stress}
\end{equation}
where the definitions of the parameters in Eq.\ \ref{critical_stress} are listed in Table\ \ref{tab:1}. Although the particle radius ultimately cancels from Eq.\ \ref{critical_stress}, it is included here so that each term is dimensionless, as in Ref. \cite{singh2007cracking}. We have included typical values for the shear modulus of both cornstarch and CaCO$_3$ taken from the literature \cite{katz2013}, assuming a crystalline form of CaCO$_3$ and a Poisson's ratio of 0.5 for cornstarch. The same values of $\phi$ and $M$ were used for all calculations. 

\begin{table}[!]\caption{Physical properties of the parameters in Eq.\ \ref{critical_stress}.}
\centering
\begin{tabular}{ccc}
     \hline\hline
     Symbol                                     &Meaning                                  &Value \\ 
     \hline
     \multirow{2}{1em}{$G$}                                        & \multirow{2}{6.5em}{particle shear modulus}                  &4 GPa (cornstarch)\\
                                               &                    &                32 GPa (CaCO$_3$)\\
    
     \multirow{2}{1em}{$R$}                                        &\multirow{2}{6.5em}{particle radius}                     &5 $\mu$m (cornstarch)\\
                                               &                      &              1 $\mu$m (CaCO$_3$)\\
      
    \multirow{3}{1em}{$\gamma$}                                  &\multirow{3}{6.5em}{surface tension}                 &72 mN/m (water)\\ 
								&			 & 				23 mN/m (IPA)\\
								&			 & 				16 mN/m (silicone oil)\\
   $M$                                      &coordination number           &5 \\
     $\phi$               &random close packing           &0.67\\
$h_c$                                   &critical thickness                  &\\
      $\sigma_ c$                                &critical stress                &\\

     \hline\hline
\label{tab:1}
\end{tabular}
\end{table}

Taking the typical values of the suspensions of cornstarch-water and CaCO$_3$-water used in the experiment (see Table\ \ref{tab:1}), we can calculate the critical thickness for cracking. Since the stress driving the primary cracks is due to capillary pressure \cite{lee2004drying}, we can assume that $\sigma_c\approx\gamma/R$. Solving for $h_c$, we obtain $h_c\approx$ 1500 $\mu$m for cornstarch, and $h_c\approx$ 400 $\mu$m for CaCO$_3$. In order to examine whether Eq.\ \ref{critical_stress} accurately predicts the critical film thickness for cracking, we dried both thin films of cornstarch-water and CaCO$_3$-water suspensions in petri dishes (Fig.\ S2). The critical film thickness for cornstarch-water suspensions was measured to be $h_c\approx$ 1180 $\mu$m, whereas for CaCO$_3$-water suspensions, $h_c\approx$ 550 $\mu$m. Both the critical film thicknesses we observed show good agreement with the predictions by Eq.\ \ref{critical_stress}, though slightly different from the predicted values. However, Eq.\ \ref{critical_stress} fails to explain the origin of secondary cracks in dried cornstarch-water suspensions. For the secondary cracks, there is no critical film thickness, and the cracks are visible in samples that are only a few particles thick, as will be discussed in Sec.\ \ref{sec:secondary_cracks_thin_chamber}, suggesting that the stress is not solely due to capillary pressure, as described in Ref. \cite{lee2004drying}. 

\subsection{Cornstarch particle deswelling drives secondary cracks}
\label{sec:particle_deswelling}
\begin{figure}[!]
\begin{center}
\includegraphics[width=3.4 in]{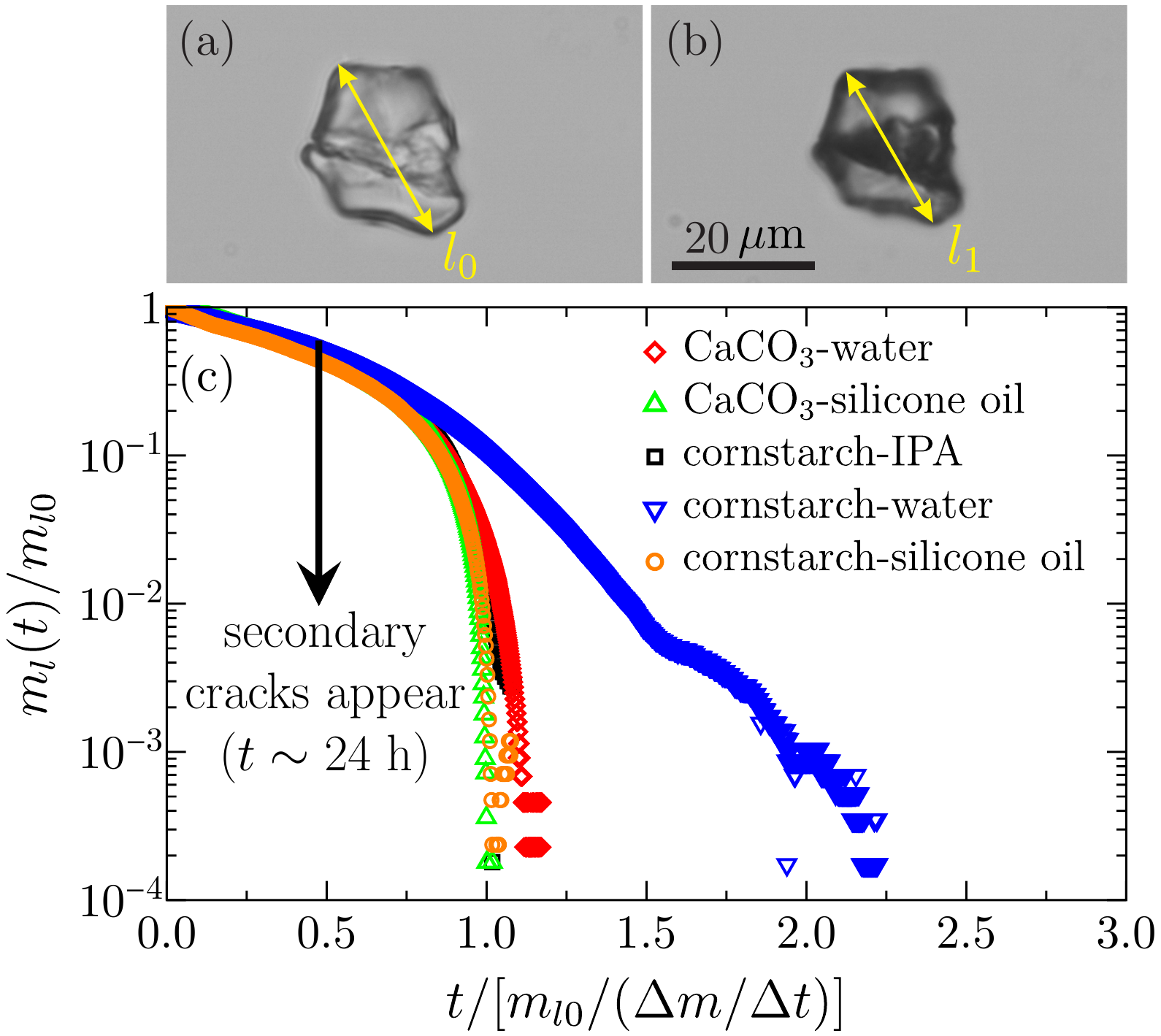}
\caption[]{Deswelling of a large, individual cornstarch particle from the wet state (a) to the fully dry state (b). The lengths indicated in the images are $l_0$ = 27 $\mu$m and $l_1$ = 24 $\mu$m. (c) Normalized evaporation rate of suspensions of cornstarch-water, -IPA and -silicone oil, and CaCO$_3$-water and -silicone oil. The symbols are defined as follows: $m_{l}(t)$ is the instantaneous mass of the liquid in the prepared suspension, $m_{l0}$ is the initial mass of the liquid, $\Delta m/\Delta t$ is the initial evaporation rate of the liquid at $t=0$.
\label{deswelling}
} 
\end{center} 
\end{figure}

Multi-scale crack patterns in dried cornstarch-water suspensions have been reported previously \cite{bohn2005hierarchicalI,bohn2005hierarchicalII,akiba2017morphometric}. The secondary cracks in cornstarch-water suspensions have been used as a model system to investigate the formation of geophysical columnar joints \cite{muller1998starch,muller1998experimental,toramaru2004columnar,goehring2008scaling,goehring2009drying,goehring2009nonequilibrium,goehring2013evolving}.
However, to the best of our knowledge, the origin of the different types of cracks is not well-understood. It has been suggested that the small-scale cracks are driven by the spatial nonuniformity of the local shrinkage of the film \cite{mizuguchi2005directional,akiba2017morphometric}, which has never been confirmed. More recently, \citet{goehring2009drying} showed that the strong separation between two distinct drying regimes dominated by liquid and vapor transport in the particle network could influence the formation of small-scale crack patterns in cornstarch-water suspensions, yet similar physics should apply in other particle networks where small-scale cracks are not observed. Consequently, the underlying mechanism for the formation of multi-scale crack patterns remains unclear. 

One of the main results of this work is that distinct polygonal crack patterns are due to distinct shrinkage mechanisms. The initial evaporation of the suspending liquid creates capillary stress at the interface, which shrinks the sample and induces stresses sufficient for cracking (primary cracks). For cornstarch in water, the secondary cracks are driven by a second shrinkage mechanism: the deswelling of the particles. We observed strong deswelling by drying swollen cornstarch particles dispersed in dilute suspensions, and the average deswelling ratio was $\approx$ 5-10\%, as shown in Fig.\ \ref{deswelling}a-b. 

In order to examine whether particle deswelling is unique for cornstarch-water suspensions, we prepared suspensions of cornstarch-water, -IPA, and -silicone oil, and suspensions of CaCO$_3$-water and -silicone oil and compared their drying kinetics. We used an electronic balance to record the instantaneous mass of the prepared suspensions during drying, and the results are shown in Fig.\ \ref{deswelling}c. The instantaneous mass of the liquid, $m_l(t)$, is normalized by the initial mass of the liquid, $m_{l0}$ (vertical axis), and the drying time $t$ is normalized by the initial evaporation rate of the liquid, $m_{l0}/(\Delta m/\Delta t)$ (horizontal axis). It can be easily seen in Fig.\ \ref{deswelling}c that the normalized drying dynamics of all of the suspensions follow the same curve, except for cornstarch and water. The drying dynamics are much slower at late times for cornstarch in water, and the sample takes more than twice as long to dry. This discrepancy suggests that the cornstarch particles are deswelling in the later drying stage, so that evaporation depends on diffusion of water out of the individual particles. Finally, the point where the drying rate of the cornstarch-water suspensions start to deviate from the other four suspensions is exactly when small-scale, secondary cracks show up during drying cornstarch-water suspensions in petri dishes ($t\sim 24 $ h, see Fig.\ \ref{hierarchical_cracks}), indicating that the secondary cracks are driven by deswelling-induced shrinkage.

\begin{figure*}[h]
\begin{center}
\includegraphics[width=6.5 in]{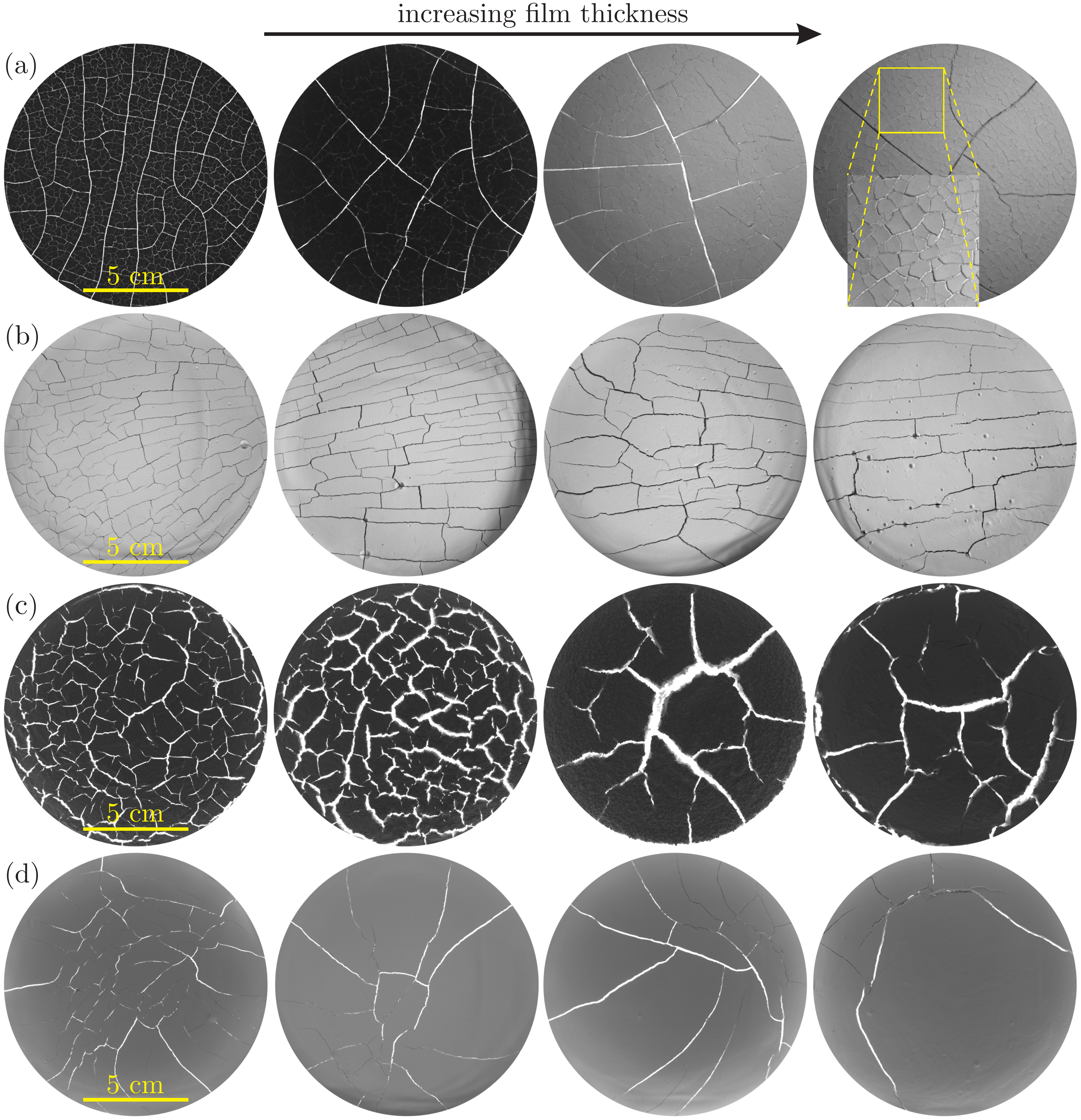}
\caption[]{Multiscale crack patterns observed in dried films of cornstarch and CaCO$_3$ particles suspended in different fluids. Panel (a) shows images of polygonal crack patterns in dried film of cornstarch in water with film thicknesses of 2, 7, 10, and 20 mm, from left to right, respectively. Panels (b), (c) and (d) show dried crack patterns of films of  CaCO$_3$ suspended in water, silicone oil, and IPA, respectively. In (b) the film thicknesses from left to right are 1.5, 2, 3, 5.5 mm; in (c) from left to right are 2.5, 4, 7.5, 9 mm; and in (d) from left to right are  3.1, 5, 5.5, 9.5 mm.
\label{allcracks}
} 
\end{center} 
\end{figure*}

\subsection{Primary cracks in different particle suspensions}
\label{sec:multi-scale cracks}

Above the critical thickness, we explored primary crack patterns in various suspensions. We dried suspensions of cornstarch and CaCO$_3$ in water, silicone oil, and IPA in petri dishes. We varied the initial volumes of the suspensions in order to obtain different film thicknesses and polygon areas. Figure\ \ref{allcracks}a shows the multi-scale cracks observed in dried cornstarch-water films, for different film thicknesses. For primary polygonal cracks, the average polygon area increases with thickness, whereas for secondary cracks, the average polygon area initially increases with film thickness and then saturates for large $h$. For CaCO$_3$-water suspensions (Fig.\ \ref{allcracks}b), the polygonal pattern is not isotropic, and has a preferred direction. This anisotropy is well-known, and is likely due to particle chain formation induced by drying \cite{akiba2018}. For CaCO$_3$ particles in both silicone oil and IPA (Fig.\ \ref{allcracks}c and \ref{allcracks}d), the cracks are distinctly different than those observed in water, nevertheless, the characteristic size of the polygons increase with thickness. This dependence will be discussed in detail in Sec.\ \ref{sec:universal_scaling}.

Although the liquids used in our experiments have different surface tensions and vapor pressures, we suspect that some of the differences in patterns are mostly due to particle-liquid interactions and surface energies. For example, we have tested the packing ability of various particle-liquid combinations (Figs.\ S3 and S4 \cite{Ma2018cracks}), and found significant differences between the same particles in different solvents. The ultimate packing fraction obtained after desiccation can potentially affect the maximum strain attainable upon drying, the modulus and tensile strength of the sample, the adhesion to the underlying substrate, and the visibility of the cracks. For example, after centrifuging prepared suspensions, we found that cornstarch particles pack significantly more densely ($\phi\approx 0.51$) in IPA than either water ($\phi\approx 0.45$) or silicone oil ($\phi\approx 0.43$), as shown in Fig.\ S3 \cite{Ma2018cracks}. In addition, both cornstarch and CaCO$_3$ pack more loosely than spherical glass beads in water (Fig.\ S4). 
\begin{figure}[!]
\begin{center}
\includegraphics[width=3.4 in]{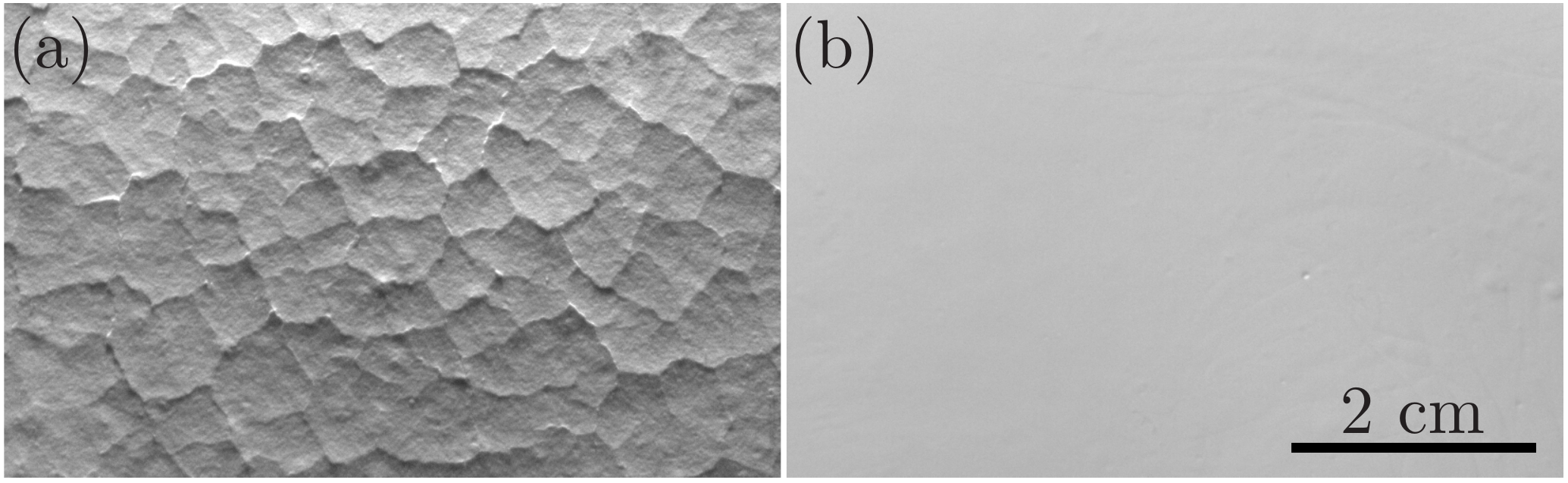}
\caption[]{Fully desiccated suspensions of cornstarch-IPA (a) and cornstarch-silicone oil (b). The two suspensions have the same initial volume (124 ml), and the same initial volume fraction ($\phi_i=40\%$). The film thicknesses of the samples in (a) and (b) are $\approx 6$ mm. The images are enhanced for better visualizations, and the scale bar applies to both images.
\label{fluids_effect}
}
\end{center} 
\end{figure}

We also observed a stark contrast in the crack patterns for cornstarch in both IPA and silicone oil as shown in Fig.\ \ref{fluids_effect}. For IPA suspensions, we only observed very fine surface cracks in thick samples ($h\gtrsim$ 1 cm), which did not penetrate more than $\approx$ 1 mm into the desiccated material (Fig.\ \ref{fluids_effect}a). We suspect that this is due to high packing density of cornstarch in IPA (Fig.\ S3 \cite{Ma2018cracks}), so that only the surface layer could obtain sufficient strain to crack upon desiccation. Suspensions of cornstarch in silicone oil did not display any visible cracks for most thicknesses used in our experiments  (Fig.\ \ref{fluids_effect}b), and only small cracks for very thick samples ($h\approx$ 2 cm). Even after full desiccation, the surface of these films looked like smooth paste  (Fig.\ \ref{fluids_effect}b), suggesting that the particles retained some sort of sticky interactions, possibly due to residual silicone oil adhered to the surface. Even weak, attractive interactions are expected to have a significant effect on granular packings and their mechanical properties for large system sizes \cite{koeze2018sticky}. This hypothesis is consistent with the low packing density of cornstarch in silicone oil (Fig.\ S3 \cite{Ma2018cracks}), suggesting that the particles may have a strong affinity for the silicone oil.

\subsection{Cornstarch-water suspensions in thin chambers}
\label{sec:secondary_cracks_thin_chamber}
\begin{figure}[!]
\begin{center}
\includegraphics[width=3.4 in]{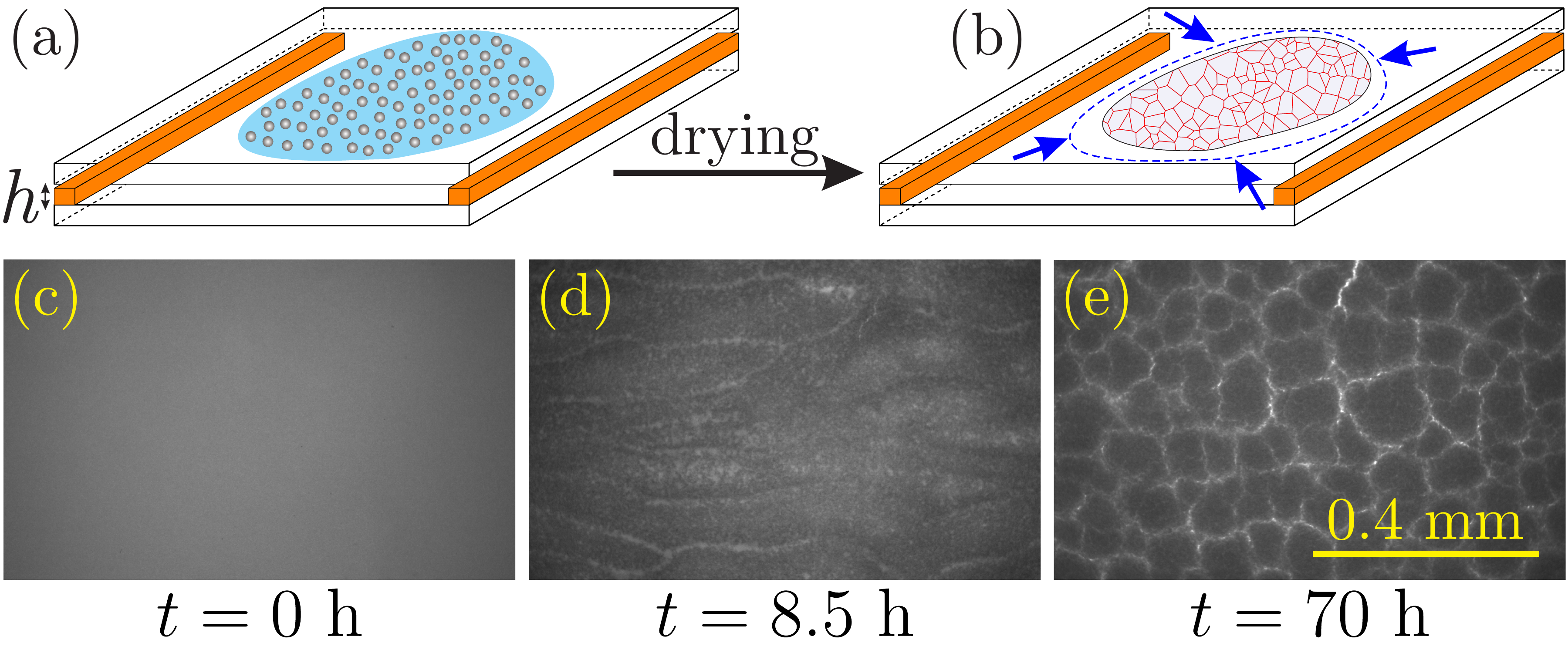}
\caption[]{ (a) Experimental setup for drying cornstarch-water suspensions in quasi-two-dimensional chambers with thickness $h$. (b) The polygonal cracks in the final dried film are indicated by the red polygons. The dashed blue line represents the profile of the initial suspension, and polygonal cracks (red) appear after the shrinkage of the initial suspension (blue arrows).
Images in (c), (d) and (e) show the drying stages: (c) the initial suspension in the thin chamber, (d) the percolation of dried regions, which appear darker in color since they scatter more light, and (e) the final polygonal crack patterns in the dried film. The scale bar applies to all of the three images. 
\label{sketch_thin_chamber}
} 
\end{center} 
\end{figure}

The small-scale secondary cracks observed in Fig.\ \ref{hierarchical_cracks}i are a unique feature of cornstarch-water suspensions, and did not show evidence of a critical thickness ($h_c$), in contrast to primary cracks. In order to explore the thickness dependence of the secondary cracks, we prepared cornstarch-water suspensions with different initial volume fractions, $\phi_i$, and then deposited the suspensions into the thin, quasi-two-dimensional chambers, as shown in Fig.\ \ref{sketch_thin_chamber}a-b. We observed two drying stages: initially a compaction front invades throughout the film; then a second drying stage ``percolates'' throughout the film with a characteristic branching pattern, leading to the formation of liquid, capillary bridges between particles.  Finally, the liquid bridges dried up followed by the formation of polygonal cracks after several days (see Fig.\ \ref{sketch_thin_chamber}c-e).  Videos showing the drying dynamics of cornstarch suspensions in thin chambers with $h=$ 750 $\mu$m (Video\_S2) and 10 $\mu$m (Video\_S3) can be found in the Supplemental Material \cite{Ma2018cracks}.

Figure \ref{dryingpattern} shows images of the polygonal cracks observed in the dried films of a cornstarch-water suspension ($\phi_i= 26\%$) in chambers with increasing $h$. Note here the thickness of the chamber is safely taken as the thickness of the dried film since the final dried film was attached to the top and bottom surfaces of the chambers, as shrinkage mostly occurred in the plane (Fig.\ \ref{sketch_thin_chamber}b). In very thin chambers ($h \simeq R$), ``dendritic'' fracture patterns are observed, as shown in Fig.\ \ref{dryingpattern}a-b. Since the thickness of the chamber is comparable to the particle size, these patterns were sensitive to the flow of the suspending liquid as evaporation occurred. As $h$ exceeds a critical value, the sensitivity to flow ceased, and regular, polygonal cracks appeared, as shown in Fig.\ \ref{dryingpattern}c-j. This trend holds true for all values of $\phi_i$ used in the experiments. 

It should be noted here that we also prepared suspensions of glass beads and CaCO$_3$ particles in thin chambers with water, however, no cracks appear during drying of the suspensions in these chambers. This supports the hypothesis that the secondary cracks are due to a distinct drying mechanism involving deswelling of the hygroscopic cornstarch particles (Sec.\ \ref{sec:particle_deswelling}). 

\begin{figure*}[t!]
\begin{center}
\includegraphics[width=1\textwidth]{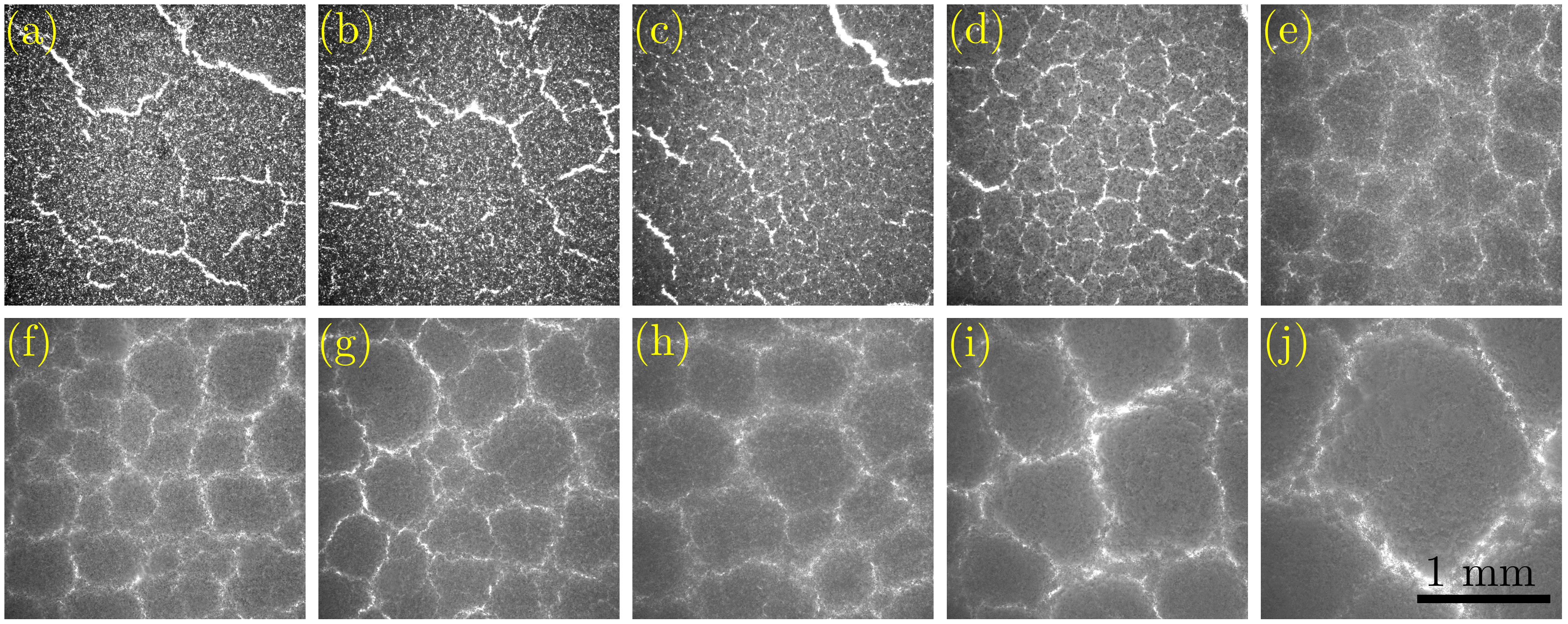}
\caption[]{Polygonal crack patterns of fully-desiccated cornstarch-water suspensions in thin chambers with initial particle volume fraction $\phi_i=26\%$. Panels (a) to (h) represent the film thicknesses of 10, 25, 50, 100, 250, 400, 500, 600, 750, and 1000 $\mu$m, respectively. The scale bar applies to all images.
\label{dryingpattern}
} 
\end{center} 
\end{figure*}
\begin{figure*}[!]
\begin{center}
\includegraphics[width=0.9\textwidth]{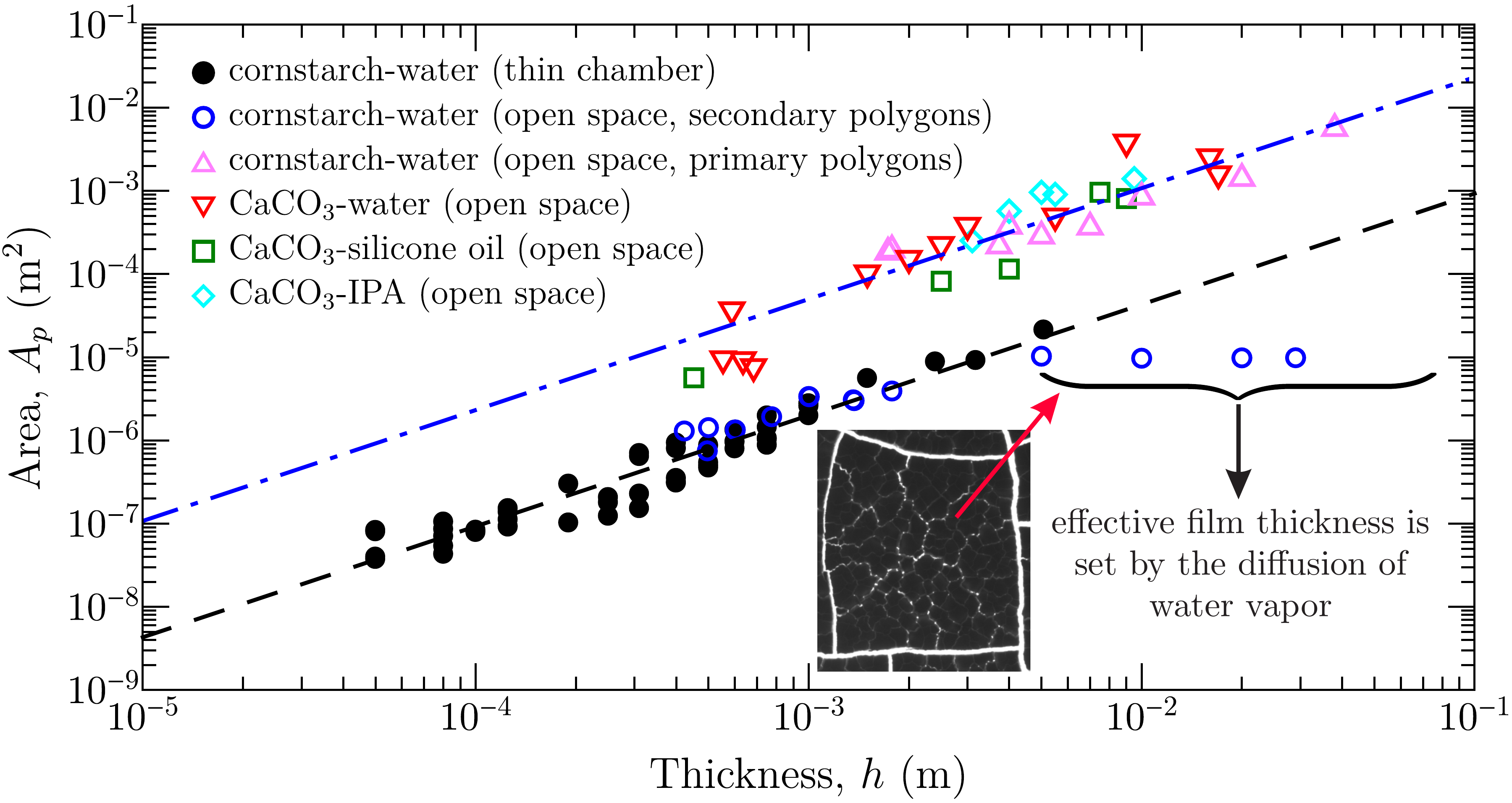}
\caption[]{Characteristic area ($A_p$) vs. the film thickness ($h$) of the multi-scale polygonal cracks observed in both thin chambers and petri dishes. The open symbols represent the data obtained in petri dishes, whereas the solid symbol represent cornstarch-water suspensions dried in thin chambers with various $\phi_i$, from Fig.\ S5 \cite{Ma2018cracks}. The open blue circles represent small-scale, secondary cracks observed in cornstarch-water suspensions dried in petri dishes. The dot-dashed blue and dashed black lines represent $A_p=\alpha h^{4/3}$, where $\alpha$ = 0.5 m$^{2/3}$ and 0.02 m$^{2/3}$, respectively.
\label{scaling_all_cracks}
} 
\end{center} 
\end{figure*}

\subsection{Universal scaling of multi-scale polygonal cracks}
\label{sec:universal_scaling}

Figures \ref{allcracks} and \ref{dryingpattern} show that the characteristic size of all observed polygonal cracks, in both petri dishes and thin chambers, increases with $h$. Although there are many ways to characterize the polygonal patterns, such as the average number of edges, or the aspect ratio, we simply measured the average area $A_p$ of the polygonal cracks. The results are shown in Fig.\ \ref{scaling_all_cracks}. The solid symbols represent cornstarch-water suspensions desiccated in thin chambers. The data were nearly independent of the initial volume fraction deposited in the chamber (Fig.\ S5 \cite{Ma2018cracks}). The open symbols represent polygons observed in petri dishes. More specifically, the open blue circles represent the secondary cracks in dried cornstarch-water suspensions in petri dishes, which overlap with the data from the thin chambers. For very thick suspensions of cornstarch and water, the area of the small-scale, secondary polygons saturated, and did not increase further. As we will show in Sec.\ \ref{effective}, the deswelling of the particles proceeds as a drying front that penetrates diffusively into the material. Thus, the effective thickness associated with the crack formation depends on the diffusion of water vapor from the film.

As mentioned previously, numerous authors have investigated the thickness dependence of the characteristic crack spacing or polygon area in desiccated suspensions \cite{groisman1994experimental,allain1995regular,komatsu1997pattern,shorlin2000alumina,lee2004drying,ma2012possible}. However, for polygonal crack networks, most of these studies cover a very limited range in thicknesses, so that a comprehensive picture of the thickness dependence of polygonal cracks is lacking. Although there is significant variation in the data from any single set of experiments, taken together, our results in Fig.\ \ref{scaling_all_cracks} strongly suggests
\begin{align}
&A_p=\alpha h^{4/3}
\end{align}
over a wide range of thickness, for different types of polygonal cracks, in different experimental geometries, and for different liquid-particle combinations. This scaling law is indicated by the dot-dashed blue and dashed black lines in Fig.\ \ref{scaling_all_cracks}. Although the prefactor, $\alpha$, is distinct for primary and secondary cracks in different materials, the data suggest that the exponent is universal. Recently, Flores \cite{flores2017mean} derived this simple scaling law using continuum elastic theory, and a balance of surface energy and elastic energy for the initiation of cracking. We will repeat this argument here since we will make small alterations to the expression for $\alpha$.

\begin{figure}[!]
\begin{center}
\includegraphics[width=3.0 in]{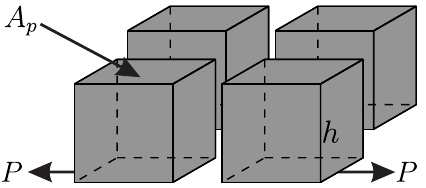}
\caption[]{Cracking in a material driven by tensile stress at the substrate. Here, $h$ is the film thickness, $A_p$ is the characteristic area of the polygonal cracks, $P$ is the tensile stress generated due to the adhesion of the film to the substrate. Adapted from Ref.\ \cite{flores2017mean}.
\label{sketch}} 
\end{center} 
\end{figure}

Figure\ \ref{sketch} shows a section of a thin film after the formation of cracks \cite{flores2017mean}. The thickness of the film is $h$, and the characteristic area of the polygons is $A_p$. The main tensile stress, $P$, acts on the bottom surface, where the film is adhered to the substrate. We assume that cracks will form when the energy cost of creating new surfaces at the sides of a polygon is equal to the elastic energy released during cracking:
\begin{align}
&\kappa\gamma \sqrt{A_p}h\sim\dfrac{V}{2E}\left<\sigma\right>^2,
\label{grif}
\end{align}
where $\gamma$ is the surface tension of the newly-created interfaces which have a typical area $\sqrt{A_p} h$, and $\kappa$ is the ratio of the perimeter to the area of the polygons. Here, $\left<\sigma\right>$ is the volume-averaged tensile stress in the film, $E$ is Young's modulus, and $V=A_p h$ is the volume of one polygon.

We can relate $\left<\sigma\right>$ to $P$ using the fact that the average stress in a volume element can be related to an integral over the forces acting on its boundaries \cite{landau1986theory}:
\begin{align}
&\left<\sigma_{ij}\right>=\dfrac{1}{2 V}\oint_S \left( P_i x_j+P_j x_i \right)\mathrm{d}S.
\label{intstress}
\end{align}
Since the main stress is applied at the substrate, Eq.\ \ref{intstress} provides the approximate scaling
\begin{align}
&\left<\sigma\right>\approx P\dfrac{\sqrt{A_p}}{h}.
\label{stressscale}
\end{align}
Although we have not rigorously evaluated the integral for a thin, adhered film, one obtains the same result, $\left<\sigma\right>\propto 1/h$, by considering a similar problem, a thin, spherical elastic shell under uniform pressure. In this case, which can be solved exactly, the tangential, tensile stress scales in the same way as Eq.\ \ref{stressscale} \cite{lautrup2011}.

Combining Eqs.\ \ref{grif} and \ref{stressscale}, we arrive at the predicted scaling relation:
\begin{align}
&A_p=\left(\dfrac{2\kappa\gamma E}{P^2}\right)^{2/3}h^{4/3}.
\label{polygon_vs_h}
\end{align}
Given the surface tension of the new, ``wet'' interfaces, $\gamma$, the prefactor, $\alpha=(2\kappa\gamma E/P^2)^{2/3}$, is determined by the modulus of the material when cracks form, and stress at the substrate generated by shrinkage, $P$. When cracks form, $P$ will essentially be the yield stress, and will be smaller than $E$ \cite{mahaut2008rheology}. Most of the polygons we observe are convex. In this case, we can estimate $\kappa$ by assuming they are regular polygons, where $\kappa$ has an analytic expression:
\begin{align}
&\kappa=2\sqrt{\dfrac{N}{\cot(\pi/N)}}.
\label{polygon_vs_h}
\end{align}
Here, $N$ is the number of sides of the polygon. For $N$ = 3, $\kappa\approx$ 4.56. As $N\rightarrow\infty$, $\kappa\rightarrow$ 3.54. Thus, in our further discussion, we will assume $\kappa\approx$ 4 for simplicity in estimating the prefactor $\alpha$. 

\begin{figure}[!]
\begin{center}
\includegraphics[width=3.4 in]{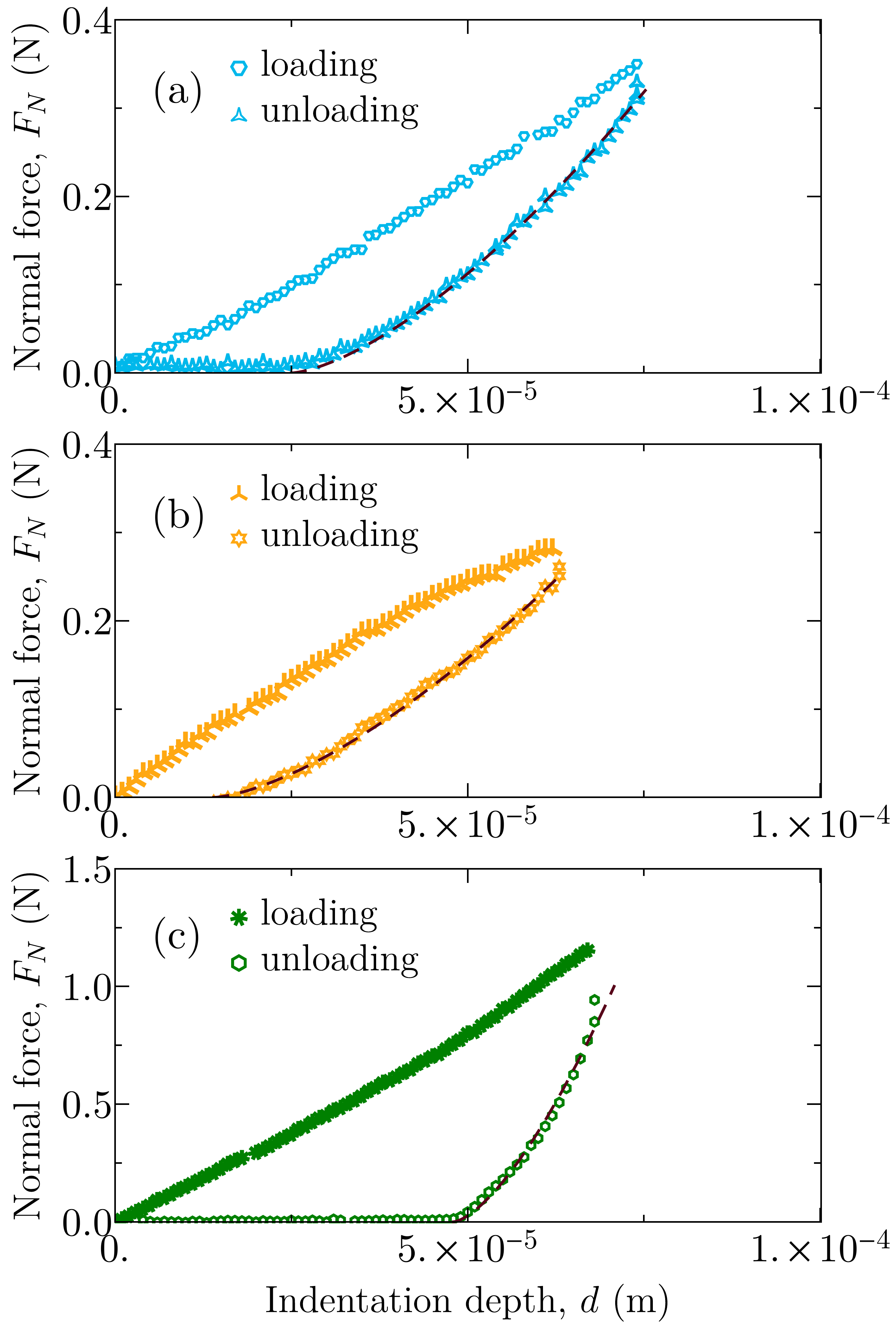}
\caption[]{Indentation load-displacement measurement of particulate suspensions at different drying stages. (a), (b) and (c) show the data of a cornstarch-water film after primary cracks appear, a cornstarch-water film when secondary cracks appear, and a CaCO$_3$-water film when cracks appear, respectively. The dashed curves represent the best fits of the data using Eq.\ \ref{loading} with corresponding modulus values $E$ = $7.3\times 10^6$ Pa (a),  $6.0\times 10^6$ Pa (b),  and $7.5\times 10^7$ Pa (c).
} 
\label{modulus}
\end{center} 
\end{figure}

In order to provide some quantitative measurement of the modulus, we prepared different mixtures at different drying stages, and used a rheometer to measure the applied normal force upon indenting the material with a stainless steel ball of radius $R_b\approx 9.5$ mm using the indentation load-displacement method \cite{oliver2004measurement}. This method assumes a linearly-elastic Hertzian contact in combination with some permanent plastic deformation. The loading-displacement relation is:
\begin{align}
\label{loading}
F_N&=\frac{4}{3} \sqrt{R_m} E^*(d-d_f)^{3/2},\\
\label{modulusrelation}
E^*&=\frac{E}{1-\nu^2},
\end{align}
where $d_f$ is the permanent depth of penetration after the indenter is fully unloaded, $E$ is Young's modulus, $\nu$ is the Poisson's ratio of the films ($\approx 0.5$). The radius $R_m=(1/R_b+1/R_h)^{-1}$, where $R_h$ is the radius of the spherical hole at the maximum indentation depth. In our experiments, $R_h\approx R_b$ and $R_m\simeq R_b/2$ since there is significant plastic deformation. Figure\ \ref{modulus} shows the load versus displacement for different particulate films when primary and secondary cracks form. Note for each tested film, the entire film thickness is at least 100 times greater than the indentation depth $d_f$. 

By fitting data to Eq.\ \ref{loading}, as indicated by the dashed curves in Fig.\ \ref{modulus}, the Young's modulus of the films fall in the range from 6-75 MPa. These values are consistent with similar modulii measured in non-Brownian dense suspensions and soils \cite{moller2007shear,mahaut2008rheology,kezdi1974,obrzud2018}. In addition, we also measured the Young's modulus of other particle-liquid combinations, and the results can be found in Fig.\ S6 \cite{Ma2018cracks}. The errors of the nonlinear fitting parameters using Eq.\ \ref{loading} range from 0.4\% to 2.0\%. The results of Fig.\ \ref{modulus}a and \ref{modulus}b are consistent with experimental observations that small-scale secondary cracks have a smaller prefactor, as shown in Fig.\ \ref{scaling_all_cracks}, although this alone does not explain the difference (0.02 m$^{2/3}$ versus 0.5 m$^{2/3}$, Fig.\ \ref{scaling_all_cracks}).

The value of the substrate stress, $P$, is more difficult to measure. The maximum adhesion to the substrate clearly affects the crack pattern, as shown in Fig.\ S1 \cite{Ma2018cracks}, so that a stronger maximum stress at the substrate decreases the polygon area, as indicated by Eq.\ \ref{polygon_vs_h}. Even for strong adhesion, shrinkage in the film will continue until $P$ is approximately the yield stress of the particle network.

For a random, close-packed particle network, the yield stress, $Y$, can be estimated as \cite{goehring2013plasticity}:
\begin{equation}
Y=\dfrac{\phi M F_\mathrm{max}}{4\pi R^2},
\label{yield_stress}
\end{equation}
where $\phi$ is the volume fraction of particles, $M$ is the coordination number, and $R$ is the particle radius (see Table\ \ref{tab:1}). $F_{\mathrm{max}}$ is the maximum force between particles. For the initiation of primary cracks, the suspension is still saturated with liquid, so we can assume that $F_{\mathrm{max}}/R^2$ can be simply estimated as the capillary pressure, $\gamma/R$. Assuming typical values of the parameters from Table\ \ref{tab:1}, Eq.\ \ref{yield_stress} gives:
\begin{equation}
Y\sim 0.27 \dfrac{\gamma}{R}.
\label{simplify_yield_stress}
\end{equation}

This result is not surprising for a wet sample if the inter-particle adhesion in the bulk liquid is small, i.e. the stress required to form a crack by pulling particles apart is of order the capillary forces holding them together. To confirm this, we have performed rheological measurements on cornstarch-water, CaCO$_3$-water, and glass beads-water suspensions with different volume fractions (Fig.\ S7 \cite{Ma2018cracks}). Although shear thickening is observed for larger volume fractions, the maximum shear stress is smaller than $\gamma/R$ \cite{brown2012thickening}, showing that capillary forces are larger than any inter-particle force in the bulk liquid.

When a crack forms, we can assume that the boundary stress will be of the same order as the yield stress, so that $P\sim Y$. Plugging the values of $E$, $\gamma$, and $R$ for the large-scale cracks in cornstarch-water samples, and assuming $\kappa\approx 4$, Eq.\ \ref{polygon_vs_h} yields $\alpha\approx 0.4$ m$^{2/3}$, which agrees well with dot-dashed blue line in Fig.\ \ref{scaling_all_cracks}. Given the variability in the modulus $E$ between different suspensions, we do not currently have a way of collapsing all the data in Fig.\ \ref{scaling_all_cracks} for the primary cracks. For wet samples, the modulus will likely depend on the surface tension, particle size, the modulus of the particles, the particle shape, and the inter-particle friction. Nevertheless, given this large parameter space, the good agreement with Eq.\ \ref{polygon_vs_h} suggests that the polygonal crack pattern can be quantitatively understood for a range of different particles and liquids, provided some knowledge of the modulus and yield stress of the suspension.

The small-scale, secondary polygonal cracks of cornstarch-water suspensions can be observed in both open petri dishes (Fig.\ \ref{hierarchical_cracks}) and thin chambers (Fig.\ \ref{dryingpattern}). This suggests that the formation of small-scale, secondary cracks does not sensitively depend on the drying geometry, and that capillary interactions are not a dominant force, in contrast to the primary cracks. Thus, local particle adhesion likely determines the yield stress of the material. As shown in Sec.\ \ref{sec:particle_deswelling}, the cornstarch particles are swollen with water, so we do not currently have a way to estimate this adhesion. In addition, the factor of $\gamma$ in the stress balance (Eq.\ \ref{grif}) would be related to the surface energies of the particle-particle adhesion \cite{goehring2013plasticity}. Since $F_{\mathrm{max}}\propto\gamma$ for adhesive forces, then $\alpha\propto (\gamma/F_{\mathrm{max}}^2)^{2/3}\propto 1/\gamma^{2/3}$. We can then conclude that this adhesion of swollen particles must be stronger than capillary interactions since the prefactor is smaller for small-scale cracks. 

\subsection{Effective film thickness for cracks in thick cornstarch-water suspensions}
\label{effective}
\begin{figure}[!]
\begin{center}
\includegraphics[width=3.4 in]{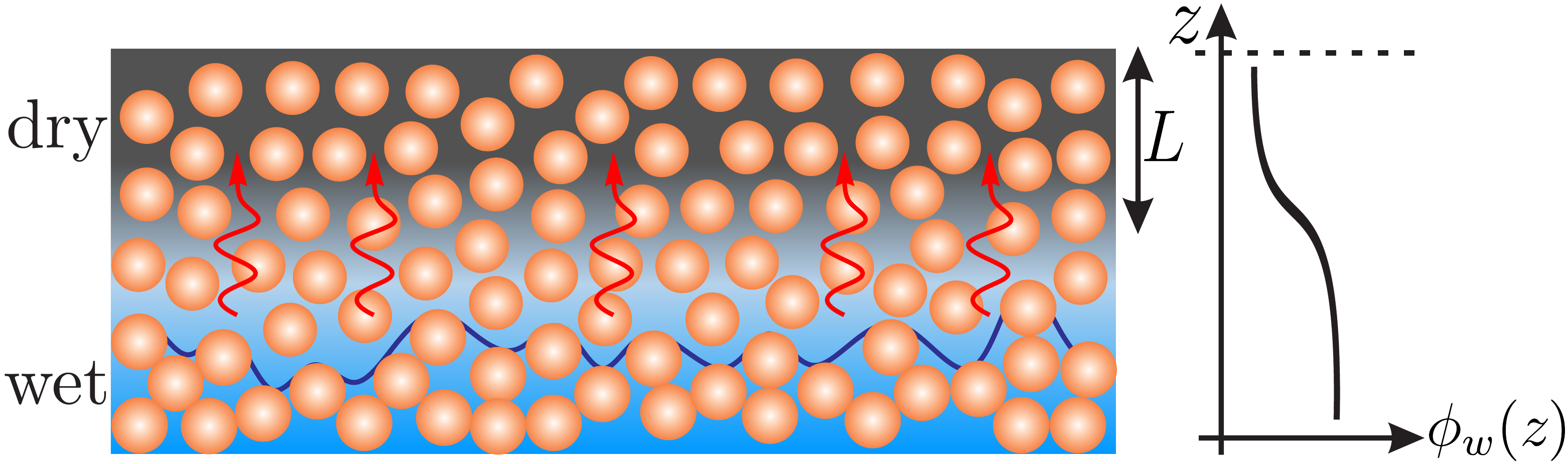}
\caption[]{The boundary between dry and wet layers during drying cornstarch-water suspensions is set by the diffusion of water vapor, as indicated by the red arrows in thick films. $L$ represents the characteristic length scale of the dry layer, i.e., the effective film thickness, and $\phi_w(z)$ is the water volume fraction along the vertical direction $z$. 
\label{boundary}
} 
\end{center} 
\end{figure}

In Fig.\ \ref{scaling_all_cracks}, we showed how the area of the small-scale, secondary polygons in cornstarch-water suspensions saturated for large values of $h$ (blue open circles). Here we show that this saturation of $A_p$ is set by the diffusion of the water vapor in the later drying stage. The individual particles will remain swollen until its environment is sufficiently dry. In an open particle network where evaporation is occurring from above, water transport is initially limited by viscosity as the liquid is pulled through the porous network according to Darcy's law. As evaporation proceeds, eventually the diffusion of water vapor through the top of the sample limits the transport. This diffusion-limited transport is likely to set a boundary between wet and dry layers, leading to an effective film thickness, $L$, as illustrated in Fig.\ \ref{boundary}. 

It has been suggested that the water vapor content can be described by a nonlinear, one-dimensional effective diffusion equation \cite{goehring2009drying,goehring2009nonequilibrium}:
\begin{equation}
\frac{\partial \phi_w}{\partial t}=\frac{\partial }{\partial z} \left[D(\phi_w) \frac{\partial \phi_w}{\partial z} \right],
\label{diffusion}
\end{equation}
where $\phi_w=\phi_w(z,t)$ is the spatio-temporal variation of the local volume fraction of water content, and $D(\phi_w)$ is local diffusivity of the water vapor, and can be expressed as \cite{pel2002analytic,goehring2009nonequilibrium}:
\begin{equation}
D(\phi_w)=\left( \int_{h}^{z} \dfrac{\partial \phi_w }{\partial t} \mathrm{d}z' \right)/\left( \dfrac{\partial \phi_w }{\partial z}  \right),
\label{diffusivity}
\end{equation}
where $0\leq z\leq h$, assuming a no-flux boundary condition at the lower boundary $z=h$. With Eq.\ \ref{diffusivity}, \citet{goehring2009nonequilibrium} measured an average $D(\phi_w) \sim 10^{-9}$ m$^2$/s with $\phi_w$ ranging from 0.1 -- 0.3 g/cm$^3$. A similar value has also been reported by M{\"u}ller \cite{muller1998starch}. In our experiment the characteristic time scale for the formation of secondary cracks in petri dishes is $T\sim$ 24 h, thus the characteristic length scale of the dry layer $L$ can be estimated as $L\sim \sqrt{TD}\sim$ 1 cm, which shows excellent agreement with the saturation thickness for small-scale cracks shown in Fig.\ \ref{scaling_all_cracks}. It should be noted that in our experiment, the cornstarch-water suspensions were dried under room temperature without introducing extra heat, so that $D(\phi_w)$ should be a bit smaller than 10$^{-9}$ m$^2$/s.

\section{Conclusions}

In summary, we experimentally investigated polygonal crack patterns in desiccated, particulate suspensions composed of various liquid and particle combinations. The thicknesses of the films ranged from $h$ = 10 $\mu$m to 4 cm. There are two major results of this work. First, the appearance of multiple, distinct length scales associated with cracks results from distinct shrinkage mechanisms during the drying process. Whereas larger, capillary-induced crack patterns occurred in many of the liquid-particle combinations, such multi-scale crack patterns only appeared in dried suspensions of cornstarch and water due to deswelling of the hygroscopic starch particles. As Fig.\ \ref{multiscale_polygonal_cracks} shows, similar multi-scale crack patterns can be observed in meter-scale, planetary terrain. This finding alone may help interpret geomorphological history from surface images, even though knowledge of the relevant material properties may not be known.

Second, the characteristic area of the polygons, for all observed cracks, is consistent with a power law scaling: $A_p=\alpha h^{4/3}$, where the prefactor is determined by a balance of surface energy ($\gamma$), film modulus ($E$), and boundary stress ($P$): $\alpha\sim(2\kappa\gamma E/P^2)^{2/3}$ \cite{flores2017mean}. The values of these parameters depend on the dominant particle-particle interaction forces at play during the initiation of cracking. By quantifying the modulus and equating $P$ with the yield stress, we are able to quantitatively predict $\alpha$ for primary cracks. We note that although this scaling law is consistent with some previous predictions \cite{komatsu1997pattern}, other authors have reported a quadratic relationship between $A_p$ and $h$ \cite{groisman1994experimental,shorlin2000alumina,leung2000pattern,leung2010criticality}. However, nearly all experimental studies report less than one order of magnitude in thickness variation. In our analysis, we have assumed that for small strains, our films can be considered homogeneous, elastic materials, and it is possible that effects such as Brownian motion, particle density fluctuations, or sticky particle interactions may explain differences observed in the literature for crack spacing and polygon area. We leave this hypothesis to future studies of other materials with different particle interactions.

Although our experiments are limited to laboratory scales, the scaling law, $A_p\propto h^{4/3}$ reproduces reasonable values for polygon areas on larger scales. For example, if we assume that polygonal cracks commonly observed in wet mud with $R\approx$ 60 $\mu$m are mainly due to capillary pressure during drying, and a typical modulus of 5 MPa \cite{kezdi1974,obrzud2018}, then Eqs.\ \ref{polygon_vs_h}, \ref{yield_stress}, and \ref{simplify_yield_stress} give $\sqrt{A_p}\approx$ 1 m for a crack depth of $h\approx$ 20 cm. For polygonal crack patterns on much larger scales, such as those show in Fig.\ \ref{multiscale_polygonal_cracks}b, we note that the polygon area may saturate due to heterogeneity in the material properties with depth. In this case it is likely that the modulus of the material is much larger, or that the stress induced during shrinkage is much smaller, in order to produce very large polygon areas. 

\section*{Acknowledgments}
We gratefully acknowledge financial support from National Science Foundation, DMR-1455086.

\bibliographystyle{apsrev4-1}
\bibliography{cracks}

\end{document}